\documentclass[aps,prd,superscriptaddress,twocolumn,showpacs]{revtex4}
\usepackage{graphicx}
\usepackage{epstopdf}
\usepackage{amsmath}
\usepackage{amsfonts}
\usepackage{amssymb}
\usepackage{latexsym}
\usepackage{verbatim}
\begin{document}
\title{Remarks on a compact Abelian Higgs model}
\author{Patricio Gaete} \email{patricio.gaete@usm.cl} 
\affiliation{Departamento de F\'{i}sica and Centro Cient\'{i}fico-Tecnol\'ogico de Valpara\'{i}so-CCTVal, Universidad T\'{e}cnica Federico Santa Mar\'{i}a, Valpara\'{i}so, Chile}
\author{Jos\'{e} A. Helay\"{e}l-Neto}\email{helayel@cbpf.br}
\affiliation{Centro Brasileiro de Pesquisas F\'{i}sicas (CBPF), Rio de Janeiro, RJ, Brasil} 
\date{\today}

\begin{abstract}
Aspects of screening and confinement are re-examined for a recently proposed compact Abelian Higgs model with a $\theta$-term. Our discussion is accomplished using the gauge-invariant but path-dependent variables formalism, which is an alternative to the Wilson loop approach. We explicitly show that the static potential profile is the sum of an effective-Yukawa and a linear potential, leading to the confinement of static external charges. We point out the central r\^ole of the parameter measuring the stiffness of the vortex lines present in the model in both the Yukawa-like and the confining sectors of the effective inter-particle potential we have computed.
\end{abstract}
\maketitle

Condensed matter physics and quantum field theory have, in the last few years, become increasingly closely meshed \cite{Hasan:2010xy,Qi:2011zya}. The physical consequences of this new physics have triggered a large body of literature. An interesting and illustrative example on this subject arises when one considers topological insulators and topological superconductors, where topological insulators have been experimentally realized in various materials. Incidentally, it is worth noting that these states are characterized by topological properties rather than usual properties such as symmetries preserved (or broken) by some order parameter. At the same time, it may be recalled that the $\theta$-term (or axion-term) has been the key ingredient of these developments, which produces non-trivial effects such as the Witten effect \cite{Witten:1979ey,Vazifeh:2010pq,Rosenberg:2010xz} and topological magneto-electric effect \cite{Li:2009tca,Essin:2008rq}.

As already expressed, the precedent states are described by a low-energy field theory which is a topological field theory. For instance, recently a topological field theory description of $(1+3)$-D topological superconductors has been discussed \cite{Qi:2012cs}. In this work, the topological superconductor is described through a topological coupling between the electromagnetic field and the superconducting phase fluctuation, in much the same way as the coupling of axions with an Abelian gauge field.

Most recently, it has been shown that a compact Abelian Higgs model with a $\theta$-term in $(1+3)$-D may be considered as an effective field theory for a topological Mott insulator \cite{Nogueira:2016sej}. Interestingly, it was also shown that this effective theory is dual to an axionic superconductor model [3], which contains both particle and vortex degrees of freedom. Curiously, it should be noted that this dual theory is analogous to a superconducting vortex strings \cite{Witten:1984eb}. Mention should be made, at this point, to an equivalent condensed matter picture from the above studies, where the vortex lines play the role of magnetic monopoles \cite{ Nogueira:2016sej}.

Let us also mention here that, in the case without matter, the previous compact Higgs model becomes a compact electrodynamics with a theta term, which is confining. It should be recalled that the existence of a phase structure for the continuum Abelian U(1) gauge theory was first obtained by including the effects due to the compactness of the U(1) group, which significantly modifies the infrared properties of the model \cite{Polyakov:1976fu}. These theories have been systematically investigated since last forty years \cite{Orland:1981ku,Kondo:1998bn}, where the crucial feature is the contribution of self-dual topological excitations. However, as we will see below, our analysis highlights that the mechanism of confinement in the model under study is not condensation of topological excitations, rather vortex lines. This is what makes the current work different from above proposals of confinement in Abelian gauge theories.

Inspired by these observations, the aim of the present work is to examine the effects of this new compact Abelian Higgs model in $(1+3)$- dimensions on a physical observable. To do this, we will work out the static potential for the present model by using the gauge-invariant but path-dependent variables formalism, along the lines of \cite{Gaete:2016grf}. In our conventions the signature of the metric is ($+1,-1,-1,-1$).

We commence our considerations with a short presentation of the model under study (a compact Abelian Higgs model with a $\theta$-term) in $(3+1)$ space-time dimensions. The model is characterized by the Lagrangian density (in Euclidean space-time) \cite{Nogueira:2016sej}:
\begin{equation}
{\cal L} =  \frac{1}{4}{\cal F}^{2}_{\mu \nu} + \frac{i{e}^2 \theta}{16 {\pi}^2}{\cal F}_{\mu \nu}{\tilde {\cal  F}_{\mu \nu}} + {\cal L}_{scalar} + \frac{1}{2 {\Omega}^2} m^2_{\mu}, \label{CAHM05}
\end{equation}
where ${m_\mu}$ is a magnetic monopole current, which is conserved. As already expressed, on the one hand, from the equivalent condensed matter picture the vortex lines play the role of magnetic monopoles such that the parameter $\Omega$ mimics the stiffness of vortex lines \cite{Nogueira} and, on the other hand, ${\Omega}^{-1}$ represents the chemical potential of monopoles. We also point out that ${\cal L}_{scalar}$ stands for the scalar field part of the Higgs model.

In writing the above we have used the notation
\begin{equation}
 {\cal F}_{\mu \nu} = F_{\mu \nu} + \frac{\pi}{\theta} {\tilde M}_{\mu \nu}, \label{CAHM051}
  \end{equation}
 and
 \begin{equation}
 {\tilde {\cal F}}_{\mu \nu} = {\tilde F}_{\mu \nu} + \frac{\pi}{\theta} {M}_{\mu \nu}, \label{CAHM052} 
 \end{equation}
 where ${M}_{\mu \nu} = {\partial}_{\mu}M_{\nu} - {\partial}_{\nu}M_{\mu}$. It may be noted here that the monopole gauge field, ${M}_{\mu}$, is given by ${M}_{\mu}(x) = \int d^4 {x}^{\prime} G(x-x^\prime){m}_{\mu}(x^{\prime})$, with the Green function $G(x) = \frac{1}{4\pi^2x^2}$ \cite{Cardy:1981fd}.

 In this context it is particularly important to recall that the subject of magnetic monopoles have a long history originating from the pioneering paper by Dirac \cite{Dirac}. The point we wish to emphasize, however, is that the fields introduced in Eqs. (\ref{CAHM051}) and (\ref{CAHM052}) are analogous to that encountered in \cite{Salam:1966bd,Cabibbo:1962td}.
  
As observed in  \cite{Nogueira:2016sej}, by introducing the auxiliary field $h_{\mu}$, equation (\ref{CAHM05}) can be brought to the form
\begin{eqnarray}
{\cal L} &=&  \frac{1}{4}({\cal F}^{2}_{\mu \nu} + {f}^{2}_{\mu \nu}) + \frac{i{e}^2 \theta}{16 {\pi}^2}{\cal F}_{\mu \nu}{\tilde {\cal  F}_{\mu \nu}}  + {\cal L}_{scalar} \nonumber\\
&+& \frac{\Omega^2}{2} ( \partial_\mu \lambda+\frac{\pi}{e}h_\mu+\frac{e\theta}{4\pi}A_\mu)^2. \label{CAHM10}
\end{eqnarray}
Notice that to get the last line it has been introduced a Lagrange multiplier, $\lambda$, in order to take into account the constraint, ${\partial}_{\mu}m_{\mu}=0$. We also have ${f}_{\mu \nu} = {\partial}_{\mu}h_{\nu} - {\partial}_{\nu}h_{\mu}$.

To proceed further, we define $z_{\mu} = \frac{\pi}{e}(h_{\mu} + \frac{e}{\pi} \partial_\mu \lambda)$. From this last definition it follows that ${f}_{\mu \nu} =  \frac{e}{\pi} {z}_{\mu \nu}=\frac{e}{\pi}({\partial}_{\mu}z_{\nu} - {\partial}_{\nu}z_{\mu})$. In this manner, we obtain the following Lagrangian density
\begin{eqnarray}
{\cal L} &=&  \frac{1}{4}{F}^{2}_{\mu \nu} + \frac{i{e}^2 \theta}{16 {\pi}^2}{F}_{\mu \nu}{\tilde {F}_{\mu \nu}}  + {\cal L}_{scalar} \nonumber\\
&+& \frac{{\Omega}^{2}e^{2}{\theta}^{2}}{32 \pi^2}A_{\mu}^2+\frac{1}{4}{z}^{2}_{\mu \nu} 
+\frac{{\Omega}^{2}e^{2}}{2{\pi}^{2}}z_{\mu}^2 \nonumber\\
&+&\frac{{\Omega}^{2}e^{2}\theta}{4{\pi}^{2}}z_{\mu}A_{\mu}. 
\label{CAHM15}
\end{eqnarray}

Next, integrating out the $z_\mu$-field induces an effective theory for the $A_{\mu}$ field, that is, 
\begin{eqnarray}
{\cal L} &=&  \frac{1}{4}{F}_{\mu \nu} \left( 1 + \frac{\frac{{\Omega}^{2}e^{2}{\theta}^{2}}{16 \pi^2}}{(-\partial^2 + \frac{{\Omega}^{2}e^{2}}{{\pi}^{2}} )} \right){F}_{\mu \nu} \nonumber\\
&+&\frac{i{e}^2 \theta}{16 {\pi}^2}{F}_{\mu \nu}{\tilde {F}_{\mu \nu}} +  
{\cal L}_{scalar}. \label{CAHM20}
\end{eqnarray}

It should be further noted that ${\cal L}_{scalar}$ is given by \cite{Nogueira}
\begin{equation}
{\cal L}_{scalar} =  \frac{\rho^2}{2} \left( \partial_\mu\phi + 2e A_\mu\right)^2. \label{CAHM25}
\end{equation}
Performing the integration over the $\phi$-field, equation (\ref{CAHM25}) reduces to
\begin{equation}
{\cal L}_{scalar} =  \frac{1}{4}{F}_{\mu \nu} \frac{4 e^2 \rho^2}{-\partial^2} {F}_{\mu \nu}.\label{CAHM30}  
\end{equation}
In summary then, the new effective Lagrangian density (in Minkowski space-time) reads 
\begin{eqnarray}
{\cal L} &=& - \frac{1}{4}{F}_{\mu \nu} \left( 1 + \frac{\frac{{\Omega}^{2}e^{2}{\theta}^{2}}{16 \pi^2}}{(\triangle + \frac{{\Omega}^{2}e^{2}}{{\pi}^{2}} )} + \frac{4e^{2} \rho^2}{\triangle}\right){F}^{\mu \nu} \nonumber\\
&+&\frac{{e}^2 \theta}{16 {\pi}^2}{F}_{\mu \nu}{\tilde {F}^{\mu \nu}}, \label{CAHM35}  
\end{eqnarray}
where $\triangle \equiv \partial_\mu \partial^\mu$.

Having described the theory under study, we can now compute the interaction energy. To this end we will calculate the expectation value of the energy operator $H$ in the physical state $ |\Phi\rangle$, which we will denote by ${\left\langle H \right\rangle _\Phi }$. 

Before going into details, we recall that this paper is aimed at studying the static potential. In such a case, we can substitute  $\Delta$ by $- {\nabla ^2}$ in equation (\ref{CAHM35}). With this remark the canonical quantization of this theory, from the Hamiltonian point of view, readily follows. The Hamiltonian analysis starts with the computation of the canonical momenta,  $\Pi^0=0$, which is the usual primary constraint equation. One can further observe that the remaining non-zero momenta are,
${\Pi ^i} = \{1- \frac{\chi^2}{(\nabla^2 - w^2)} - \frac{\varpi^2}{\nabla^2}\}E^i - \frac{\theta}{4\pi^2}B^i$, where $\chi^2=\frac{\Omega^2 e^2 \theta^2}{16 \pi^2}$, $w^2 = \frac{\Omega^2e^2}{\pi^2}$ and $\varpi^2 = 16 e^2 \rho^2$.  Thus, the canonical Hamiltonian corresponding to (\ref{CAHM35}) is given by
\begin{eqnarray}
H_{C} &=& \int d^3 x\left[ - A_0\partial_i\Pi^i +\frac{1}{2}\Pi^i \Xi^{-1}\Pi^i + \frac{1}{2}B^i \Xi B^i\right] \nonumber\\
&+& \int d^3 x\left[\frac{1}{2}B^i \Xi^{-1}\Pi^i + \frac{3\theta^2}{128\pi^2}B^i \Xi^{-1} B^i\right],  \label{CAHM40}  
\end{eqnarray}
where $\Xi = \{1- \frac{\chi^2}{(\nabla^2 - w^2)} - \frac{\varpi^2}{\nabla^2}\}$.

Time conservation of the primary constraint, $\Pi^{0}$ yields a secondary constraint $\Gamma _1  = \partial _i \Pi ^i  = 0$. The preservation of $\Gamma _1$ for all times does not 
give rise to any further constraints. The extended Hamiltonian that generates translations in time is thus given by $H = H_{C} + \int d^3x\{c_{0}(x)\Pi_{0}(x) + c_{1}(x)\Gamma_{1}(x)\}$, where $c_{0}(x)$ and $c_{1}(x)$ are  Lagrange multipliers reflecting the gauge invariance of the theory. Since $\Pi_{0}=0$ for all time and $\dot {A}_{0}(x) = \left[A_{0}(x), H\right] = c_{0}(x)$, which is completely arbitrary, we discard $A_{0}$ and $\Pi_{0}$ because they add nothing to the description of the system. Then, the Hamiltonian takes the form
\begin{eqnarray}
H &=& \int d^3 x\left[c(x) \partial_i\Pi^i +\frac{1}{2}\Pi^i \Xi^{-1}\Pi^i + \frac{1}{2}B^i \Xi B^i\right] \nonumber\\
&+& \int d^3 x\left[\frac{1}{2}B^i \Xi^{-1}\Pi^i + \frac{3\theta^2}{128\pi^2}B^i \Xi^{-1} B^i\right], 
\end{eqnarray} 
\label{CAHM45}  
where $c(x)$ is a new arbitrary Lagrange multiplier.

It is important to emphasize that the existence of this arbitrary quantity $c(x)$ is unwanted since we have no way of giving it a meaning in a quantum theory. In view of this situation, we shall introduce a gauge condition such that the full set of constraints becomes second class. A particularly helpful choice is \cite{Gaete:1997eg} 
\begin{equation}
\Gamma _2 \left( x \right) \equiv \int\limits_{C_{\zeta x} } {dz^\nu } A_\nu
\left( z \right) \equiv \int\limits_0^1 {d\lambda x^i } A_i \left( {\lambda
x } \right) = 0,  \label{CAHM50}
\end{equation}
where $\lambda$ $(0\leq \lambda\leq1)$ is the parameter describing the
spacelike straight path $x^i = \zeta ^i + \lambda \left( {x - \zeta}
\right)^i $ , and $\zeta $ is a fixed point (reference point). There is no
essential loss of generality if we restrict our considerations to $\zeta
^i=0 $. We thus obtain the only nonvanishing equal-time Dirac bracket  
\begin{eqnarray}
\left\{ {A_i \left( x \right),\Pi ^j \left( y \right)} \right\}^ * &=& \delta{\
_i^j} \delta ^{\left( 3 \right)} \left( {x - y} \right) \nonumber\\
&-& \partial _i^x
\int\limits_0^1 {d\lambda x^j } \delta ^{\left( 3 \right)} \left( {\lambda x
- y} \right).  \label{CAHM55}
\end{eqnarray}

We are now equipped to obtain the corresponding interaction energy between
pointlike sources in the model under consideration, where a fermion is
localized at $\mathbf{y}\prime$ and an antifermion at $\mathbf{y}$. Let us start by observing
that $\left\langle H \right\rangle _\Phi$ reads 
\begin{eqnarray}
\left\langle H \right\rangle _\Phi &=& \left\langle \Phi \right|\int {d^3 x} 
\Biggl[\frac{1}{2}\Pi ^i\left( {\frac{{\nabla ^2  - w^2 }}{{\left[\nabla ^2 - \epsilon ^2  \right] +\frac{w^2\varpi^2}{\nabla^2}}}} \right)\Pi ^i \nonumber\\
&+&\frac{1}{2}B ^i \left( {\frac{{\left[\nabla ^2 - \epsilon ^2  \right] +\frac{w^2\varpi^2}{\nabla^2}}}{{
\nabla ^2 - w ^2 }}} \right)B^i   \nonumber\\
&+& \frac{\theta}{4\pi^2} B_i\left( {\frac{{\nabla ^2  - w^2 }}{{\left[\nabla ^2 - \epsilon ^2  \right] +\frac{w^2\varpi^2}{\nabla^2}}}} \right)\Pi^i \nonumber\\
&+& \frac{3\theta^2}{128\pi^2}B^{i}\left( {\frac{{\nabla ^2  - w^2 }}{{\left[\nabla ^2 - \epsilon ^2  \right] +\frac{w^2\varpi^2}{\nabla^2}}}} \right)B^{i}\Biggr]|\Phi\rangle,  \nonumber\\
\label{CAHM60}
\end{eqnarray}
where $\epsilon^2 \equiv (w^2 + \chi^2 + \varpi^2)$.

In this case, it is also opportune to recall that the physical state can be written as \cite{Dirac:1955uv}
\begin{eqnarray}
\left| \Phi \right\rangle &\equiv& \left| {\overline \Psi \left( \mathbf{y }
\right)\Psi \left( \mathbf{y}\prime \right)} \right\rangle \nonumber\\
&=& \overline \psi
\left( \mathbf{y }\right)\exp \left( {iq\int\limits_{\mathbf{y}\prime}^{ 
\mathbf{y}} {dz^i } A_i \left( z \right)} \right)\psi \left(\mathbf{y}\prime
\right)\left| 0 \right\rangle,  \label{CAHM65}
\end{eqnarray}
where $\left| 0 \right\rangle$ is the physical vacuum state and the line
integral appearing in the above expression is along a spacelike path
starting at $\mathbf{y}\prime$ and ending at $\mathbf{y}$, on a fixed time
slice. As a consequence of this the fermion fields are now dressed by a cloud
of gauge fields.

Making use of the previous Hamiltonian structure we then easily verify that 
\begin{eqnarray}
\Pi _i \left( x \right)\left| {\overline \Psi \left( \mathbf{y }\right)\Psi
\left( {\mathbf{y}^ \prime } \right)} \right\rangle &=& \overline \Psi \left( 
\mathbf{y }\right)\Psi \left( {\mathbf{y}^ \prime } \right)\Pi _i \left( x
\right)\left| 0 \right\rangle \nonumber\\
&+& q\int_ {\mathbf{y}}^{\mathbf{y}^ \prime } {\
dz_i \delta ^{\left( 3 \right)} \left( \mathbf{z - x} \right)} \left| \Phi
\right\rangle.  \nonumber\\
\label{CAHM70}
\end{eqnarray}

It follows from the above equation that 
 \begin{equation}
\left\langle H \right\rangle _\Phi = \left\langle H \right\rangle _0 +
\left\langle H \right\rangle _\Phi ^{\left( 1 \right)} + \left\langle H
\right\rangle _\Phi ^{\left( 2 \right)},  \label{CAHM75}
\end{equation}
where $\left\langle H \right\rangle _0 = \left\langle 0 \right|H\left| 0
\right\rangle$, and the $\left\langle H \right\rangle _\Phi ^{\left( 1
\right)}$ and $\left\langle H \right\rangle _\Phi ^{\left( 2 \right)}$ terms
are given by 
\begin{eqnarray}
\left\langle H \right\rangle _\Phi ^{\left( 1 \right)} &=&- \frac{{{\omega ^2}{\varpi ^2}}}{{2\left( {M_1^2 - M_2^2} \right)}} \nonumber\\
&\times&\left\langle \Phi \right|\int {d^3 x} \Pi _i \left[ \frac{1}{M_2^2}  \frac{\nabla ^2}{(\nabla ^2 -
M_1^2)}\right] \Pi ^i \left| \Phi \right\rangle \nonumber\\
&+&\frac{{{\omega ^2}{\varpi ^2}}}{{2\left( {M_1^2 - M_2^2} \right)}} \nonumber\\
&\times&\left\langle \Phi \right|\int {d^3 x} \Pi _i \left[ \frac{1}{M_1^2}  \frac{\nabla ^2}{(\nabla ^2 -
M_2^2)}\right] \Pi ^i \left| \Phi \right\rangle, \nonumber\\
\label{CAHM80}
\end{eqnarray}
and
\begin{eqnarray}
\left\langle H \right\rangle _\Phi ^{\left( 2 \right)} &=&\frac{{{\omega ^2}{\varpi ^2}}}{{2\left( {M_1^2 - M_2^2} \right)}}   \nonumber\\
&\times&\left\langle \Phi \right|\int {d^3 x} \Pi _i \left[ \frac{1}{M_2^2}  \frac{1}{(\nabla ^2 -
M_1^2)}\right] \Pi ^i \left| \Phi \right\rangle \nonumber\\
&-& \frac{{{\omega ^2}{\varpi ^2}}}{{2\left( {M_1^2 - M_2^2} \right)}}  \nonumber\\
&\times&\left\langle \Phi \right|\int {d^3 x} \Pi _i \left[\frac{1}{M_1^2}  \frac{1}{(\nabla ^2 -
M_2^2)}\right] \Pi ^i \left| \Phi \right\rangle, \nonumber\\
\label{CAHM85}
\end{eqnarray}
where $M_1^2 = \frac{1}{2}\left[ \epsilon^2 + \sqrt{\epsilon^4 - 4M^4}\right]$, $M_2^2 = \frac{1}{2}\left[ \epsilon^2 - \sqrt{\epsilon^4 - 4M^4}\right]$, ${M_1} \ge {M_2} \ge 0$. Whereas $M^4=w^2\varpi^2$.

Consequently, the static potential profile for two opposite charges
located at $\mathbf{y}$ and $\mathbf{y^{\prime }}$ then reads
\begin{eqnarray}
V &=& - \frac{{{q^2}}}{{4\pi }}\frac{{{\omega ^2}{\varpi ^2}}}{{\left( {M_1^2 - M_2^2} \right)}}
\left\{ {\frac {1}{M_2^2}\frac{{e^{ - M_1 L} }}{L} -\frac {1}{M_1^2}\frac{{e^{ - M_2 L} }}{L}}\right\}  \nonumber \\
&+&\frac{{{q^2}}}{{8\pi }}\frac{{{\omega ^2}{\varpi ^2}}}{{\left( {M_1^2 - M_2^2} \right)}} \nonumber\\
&\times&\left\{ {\frac{1}{M_2^2}\ln \left( {1 + \frac{{\Lambda ^2 }}{{M_1^2 }}}
\right) - \frac{1}{M_1^2} \ln \left( {1 + \frac{{\Lambda ^2 
}}{{M_2^2 }}} \right)} \right\}L,  \nonumber\\
\label{CAHM90}
\end{eqnarray}
where $\Lambda$ is an ultraviolet cutoff and $|\mathbf{y}-\mathbf{y}^{\prime }|\equiv L$. From this last expression it follows that the effect of including vortex lines (magnetic monopoles) is a linear potential, leading to the confinement of static charges. We also call attention to the fact that the present cutoff arises when manipulating the ultraviolet divergent integral (\ref{CAHM85}). Given this situation, it will be useful to give a meaning to the cutoff $\Lambda$. In this case, we first observe that our effective theory for the electromagnetic field is an effective description that arises under integration over the $\Lambda$-field, whose excitations are massive (${k^2} = M_1^2$ and ${k^2} = M_2^2$). Evidently, ${\raise0.7ex\hbox{$1$} \!\mathord{\left/{\vphantom {1 {{M_1}}}}\right.\kern-\nulldelimiterspace}\!\lower0.7ex\hbox{${{M_1}}$}}$ and 
${\raise0.7ex\hbox{$1$} \!\mathord{\left/{\vphantom {1 {{M_1}}}}\right.\kern-\nulldelimiterspace}
\!\lower0.7ex\hbox{${{M_2}}$}}$, are the Compton wavelengths of these excitations, which define a correlation distance. We thus see that physics at distances of the order or lower ${\raise0.7ex\hbox{$1$} \!\mathord{\left/{\vphantom {1 {{M_1}}}}\right.\kern-\nulldelimiterspace}
\!\lower0.7ex\hbox{${{M_2}}$}}$ must take into account a microscopic description of vortex lines fields. In other words, if we work with energies of the order or higher than $M_{2}$, our effective theory with the integrated effects of $\Lambda$ is no longer sensible. Accordingly, we identify $\Lambda$ with $M_{1}$.
Thus, finally we end up with
\begin{eqnarray}
V &=& - \frac{{{q^2}}}{{4\pi }}\frac{{{\omega ^2}{\varpi ^2}}}{{\left( {M_1^2 - M_2^2} \right)}}
\left\{ {\frac {1}{M_2^2}\frac{{e^{ - M_1 L} }}{L} -\frac {1}{M_1^2}\frac{{e^{ - M_2 L} }}{L}}\right\}  \nonumber \\
&+&\frac{{{q^2}}}{{8\pi }}\frac{{{\omega ^2}{\varpi ^2}}}{{\left( {M_1^2 - M_2^2} \right)}} \nonumber\\
&\times&\left\{ {\frac{1}{M_2^2}\ln \left( {2}
\right) - \frac{1}{M_1^2} \ln \left( {1 + \frac{{M_1^2 
}}{{M_2^2 }}} \right)} \right\}L,  \nonumber\\
\label{CAHM95}
\end{eqnarray}

Since the inter-particle potential above describes an effective physics below the cutoff $\Lambda$ , and the mass $M_1$ has been identified with $\Lambda$, we are not allowed to consider the limit of the potential in
the limit of very high stiffness, when the magnetic monopole current -current interaction becomes negligible. This limit would correspond to an infinite value of $M_1$, but this is not possible for $M_1$ is actually the
cutoff we adopt. So, our results hold only for finite values of the stiffness parameter, $\Omega$.

Another interesting observation is that the $\rho \to 0$ limit is not always smooth, that is, from eq.(\ref{CAHM95}) it is evident that in such a case the interaction energy vanishes. This might seem unexpected since that in such a limit we should recover the known result of compact electrodynamics with a $\theta$-term. The reason for this is very simple, the algebraic structure that leads to static potential is quite different. Thus we observe that the limit  $\rho \to 0$ must be taken in eq.  (\ref{CAHM35}). We thus find that
\begin{equation}
{\cal L} = -\frac{1}{4}F_{\mu\nu}\left[ \frac{\nabla^2-w^2(1+\frac{\theta^2}{16})}{\nabla^2 - w^2} \right]F^{\mu\nu} + \frac{{e}^2 \theta}{16 {\pi}^2}{F}_{\mu \nu}{\tilde {F}^{\mu \nu}}.
\label{CAHM100}
\end{equation}
From this expression it follows that the static potential profile is analogous to that encountered in equation (\ref{CAHM90}). In addition, one can easily see that in the $\theta \to 0$, we obtain the Coulomb potential. However, the limit $w \to 0$ requires a special attention. Let us be more specific. If we consider $w \to 0$
then this means $\Omega \to 0$ (vanishing stiffness). If also matter is absent, then we have the problem of both $M_1$ and $M_2$ blowing up. Therefore, in such a situation, we cannot use we have derived to get to eq. (\ref{CAHM90}). The right procedure to follow is to go back to our Lagrangian (\ref{CAHM35}) and so it can be readily shown that the Coulomb behavior is restored.

In summary, we have studied the recently proposed compact Abelian Higgs model with a $\theta$-term \cite{Nogueira:2016sej} from a different perspective. Our discussion has been carried out by using the gauge-invariant but path-dependent variables formalism, which is an alternative to the Wilson loop approach. Once again we
have exploited a correct identification of field degrees of freedom with observable quantities. It was explicitly shown that the static potential profile is the sum of an effective-Yukawa and a linear potential, leading to the confinement of static external charges. Finally, we would like to recall that the above static potential profile is analogous to that encountered in a theory of antisymmetric tensor fields that results from the condensation of topological defects as a consequence of the Julia-Thoulouse mechanism \cite{Gaete:2004dn}. In fact, this mechanism is a condensation process dual to the Higgs mechanism \cite{Quevedo:1996uu}, which describes the electromagnetic behavior of antisymmetric tensors in the presence of magnetic-branes (topological defects) that condensate due to thermal and quantum fluctuations.

One of us (P. G.) was partially supported by Fondecyt (Chile) grant 1180178 and by Proyecto Basal FB0821.

\end{document}